\documentclass[journal=jacsat,manuscript=article]{achemso}
\usepackage[utf8]{inputenc}
\usepackage{textcomp}
\usepackage{graphicx}
\usepackage{mhchem}
\usepackage{siunitx}
\DeclareSIUnit\au{a.u.}
\usepackage{amsmath}
\usepackage{upgreek}
\usepackage{hyperref}
\usepackage{amssymb}
\usepackage{tabularx}
\usepackage{booktabs}
\usepackage{epstopdf}
\AppendGraphicsExtensions{.tiff}
\usepackage{comment}
\allowdisplaybreaks

\title{Robust Tensor Hypercontraction of the Particle-Particle Ladder Term in Equation-of-Motion Coupled Cluster Theory}

\author{Avdhoot Datar}
\email{adatar@smu.edu}
\affiliation{Department of Chemistry, Southern Methodist University, Dallas, TX 75275, USA}

\author{Devin A. Matthews}
\email{damatthews@smu.edu}
\affiliation{Department of Chemistry, Southern Methodist University, Dallas, TX 75275, USA}

\abbreviations{CCSD, EOM-CCSD, PPL, LS-THC, LS-PTHC, R-LS-THC}

\begin{document}

\begin{tocentry}
\includegraphics[scale=0.25]{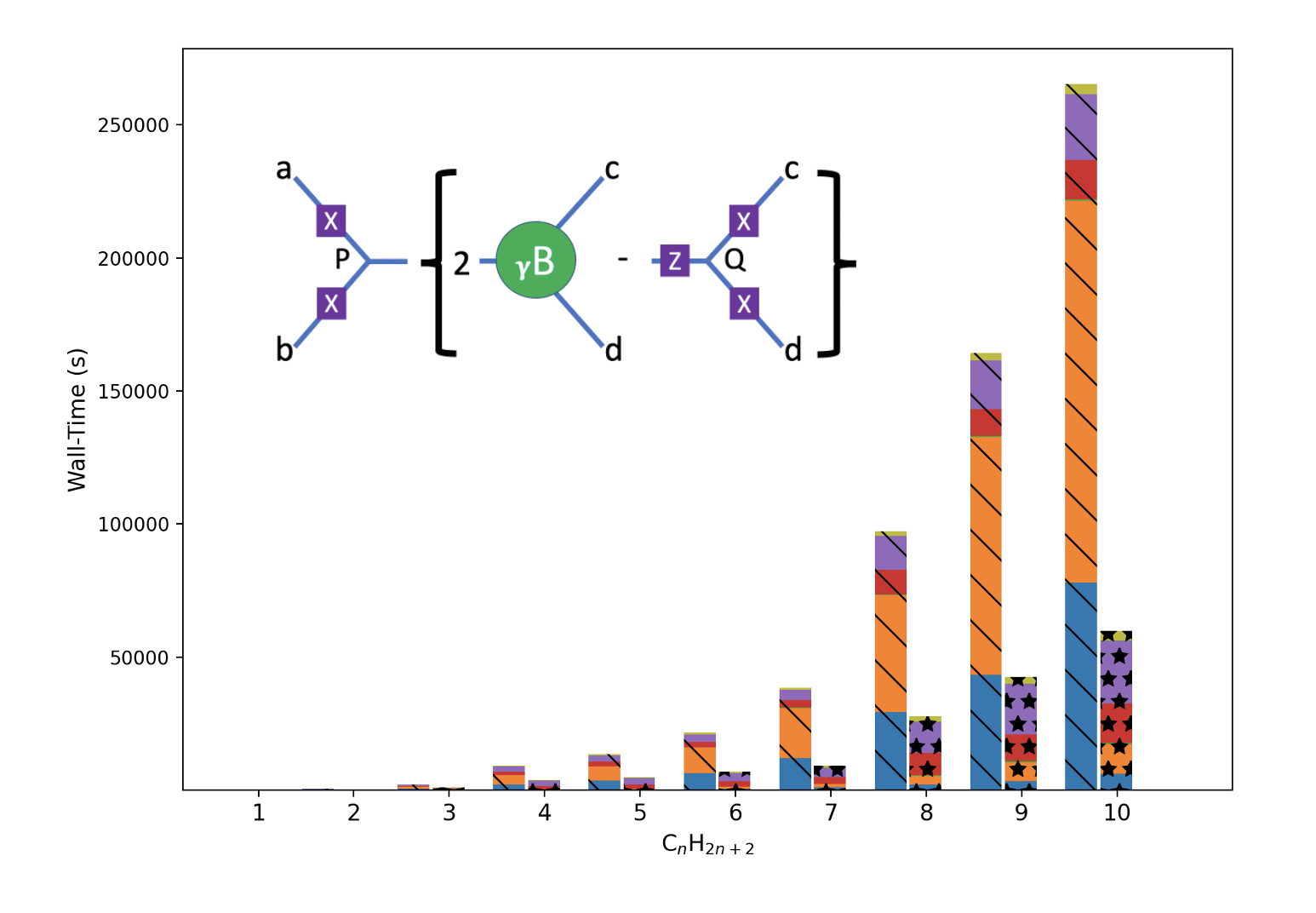}
\end{tocentry}

\begin{abstract}
One method of representing a high-rank tensor as a (hyper-)product of lower-rank tensors is the tensor hypercontraction (THC) method of Hohenstein et al. This strategy has been found to be useful for reducing the polynomial scaling of coupled-cluster methods by representation of a four dimensional tensor of electron-repulsion integrals in terms of five two-dimensional matrices. Pierce et al. have already shown that the application of a robust form of THC to the particle-particle ladder term (PPL) reduces the cost of this term in couple-cluster singles and doubles (CCSD) from $\mathcal{O} (N^6)$ to $\mathcal{O} (N^5)$ with negligible errors in energy with respect to the density-fitted variant. In this work we have implemented the least-squares variant of THC (LS-THC) which does not require a non-linear tensor factorization, including the robust form (R-LS-THC), for the calculation of the excitation and electron attachment energies using equation-of-motion coupled cluster methods EOMEE-CCSD and EOMEA-CCSD, respectively. We have benchmarked the effect of  the R-LS-THC-PPL approximation on excitation energies using the comprehensive QUEST database and the accuracy of electron attachment energies using the NAB22 database. We find that errors on the order of 1 meV are achievable with a reduction in total calculation time of approximately $5\times$.
\end{abstract}

\section{Introduction}
Coupled cluster (CC) methods provide a rigorous theoretical framework for calculating accurate molecular energies and properties for small molecules. However, the steep polynomial scaling of computational cost of these methods is an impediment for applying them to larger molecules. Many numerical approximations have been proposed to reduce the cost of CC and related electronic structure methods, such as: Cholesky decomposition (CD),\cite{CD-1965,CD-1970,CD-1977,CD-2019} resolution-of-the-identity (RI) or density fitting approximation (DF),\cite{RI-1993,RI-Pulay-1993,RI-1996,RI-1999,RI-2003} pseudospectral (PS) approach,\cite{PS-HF-1985,PS-1986,PS-1990,PS-1990-Ringnalda,PS-1991,PS-1992,PS-1993,PS-1994,PS-1995,PS-2008} fast multipole method (FMM),\cite{FMM1994,FMM1996,FMM2006} the CANDECOMP/PARAFAC (CP, also known as canonical polyadic) decomposition,\cite{CP-2007,CP-2009,CP-CCD-2013,CP-2016,CP-MP2-2011,CP2013,CP-PCC-2017} and the tensor hypercontraction (THC) approach.\cite{THC-2012,THC-2013,THC-2017,THC-RRCCSD-2019,THC2-2012,THC3-2012,THCCC2-2013,THCCCSD-2014} While the key feature in these various numerical approximations is decomposing the four-index electron repulsion integral (ERI) tensor into smaller objects (three-index tensors, matrices, etc.), these approximations can be largely divided into two groups: one based on a factorization in terms of auxiliary basis functions and the other based on a representation of ERIs using a set of physically motivated grid points.\footnote{Note that mathematically, the use of grid points for ERI approximation can often be interpreted as an integral over an auxiliary basis ``function" defined in terms of distributions such as the Dirac distribution.} Amongst these, the THC approach\cite{THC-2012} and in particular, its least-squares variant (LS-THC),\cite{THC2-2012} is quite attractive as it provides a systematic approach for achieving a chosen level of accuracy in fitting the ERI tensor, and an efficient non-iterative procedure for determining the fit parameters. Controllable accuracy can be achieved either by selecting appropriate starting grids,\cite{THC-grids} or pruning the starting grid by removing near-linear dependencies in the grid metric matrix.\cite{Matthews-grid-prune}

In coupled cluster with single and double excitations (CCSD) and related methods, the particle-particle ladder (PPL) term is the most expensive single contribution, often representing a majority of the total computational time. The relative cost of the PPL term also increases with increasing basis set size given that it scales as $\mathcal{O}(N_v^4 N_o^2)$ while other leading-order terms scale as $\mathcal{O}(N_v^3 N_o^3)$ or $\mathcal{O}(N_v^2 N_o^4)$, where $N_v$ ($N_o$) is the number of virtual (occupied) molecular orbitals. Parrish et al. have shown that LS-THC can be employed to greatly reduce the cost CCSD by factorizing the PPL term alone.\cite{THC-PPL-2014} A partial LS-THC (LS-PTHC) factorization was also explored and found to result in lower errors for the PPL term.\cite{THC-PPL-2014} Recently, the relationship between DF, LS-PTHC, and LS-THC forms of ERI factorization were used to derive a robust approximation to LS-THC-type methods, leading to significant error cancellations in fitting the ERI tensor and impressively small energy errors when applied to the PPL term of CCSD.\cite{THC-PPL-2021} In that work, a non-linear CP decomposition of the density fitting factors was used to arrive at THC and PTHC factorization forms, leading to the designation rCP-DF.

The equation-of-motion excitation energy coupled cluster framework allows for calculating excited states (EOMEE-CC), electron attachment energies (EOMEA-CC), and other related energies via linear excitation operators applied to the CC ground state wavefunction.\cite{EOM1,EOM1989,EOM2,EOM3,Nooijen1995eomea} Determination of the (vertical) excitation energy involves the solution of an eigenvalue equation via iterative application of the CC-transformed Hamiltonian to the EOMEE-CC wavefunction. As in the ground state CCSD calculation, the most expensive term in EOMEE-CCSD (as well as related methods which utilize approximate ground states, such as EOMEA-MBT2\cite{EOM-MBPT2}) methods is the PPL term. In a recent study, it was found that an efficient DF approximation, in particular applied to the PPL term, within EOM-CCSD greatly reduces the computational time.\cite{PPL-DF-2022} 

In this work, we have implemented the LS-THC, partial variant of THC (LS-PTHC), and R-LS-THC methods for the PPL term (both ground and excited state equations) within the EOMEE-CCSD and EOMEA-CCSD methods. In the following sections we discuss the mathematical details of these THC methods, as well as the performance and accuracy of the THC methods as compared to the standard DF approach.

\section{Theoretical Methods}

In this work the following notation is used: the letters $pqrs$ denote arbitrary molecular orbitals (MOs), while $ijklmn$ ($abcdef$) denote occupied (virtual) MOs. The letters $JKL$ are used to denote density fitting auxiliary basis functions, and LS-THC grid points are indicated by the letters $RS$. 

\begin{figure}
\includegraphics[scale=0.45]{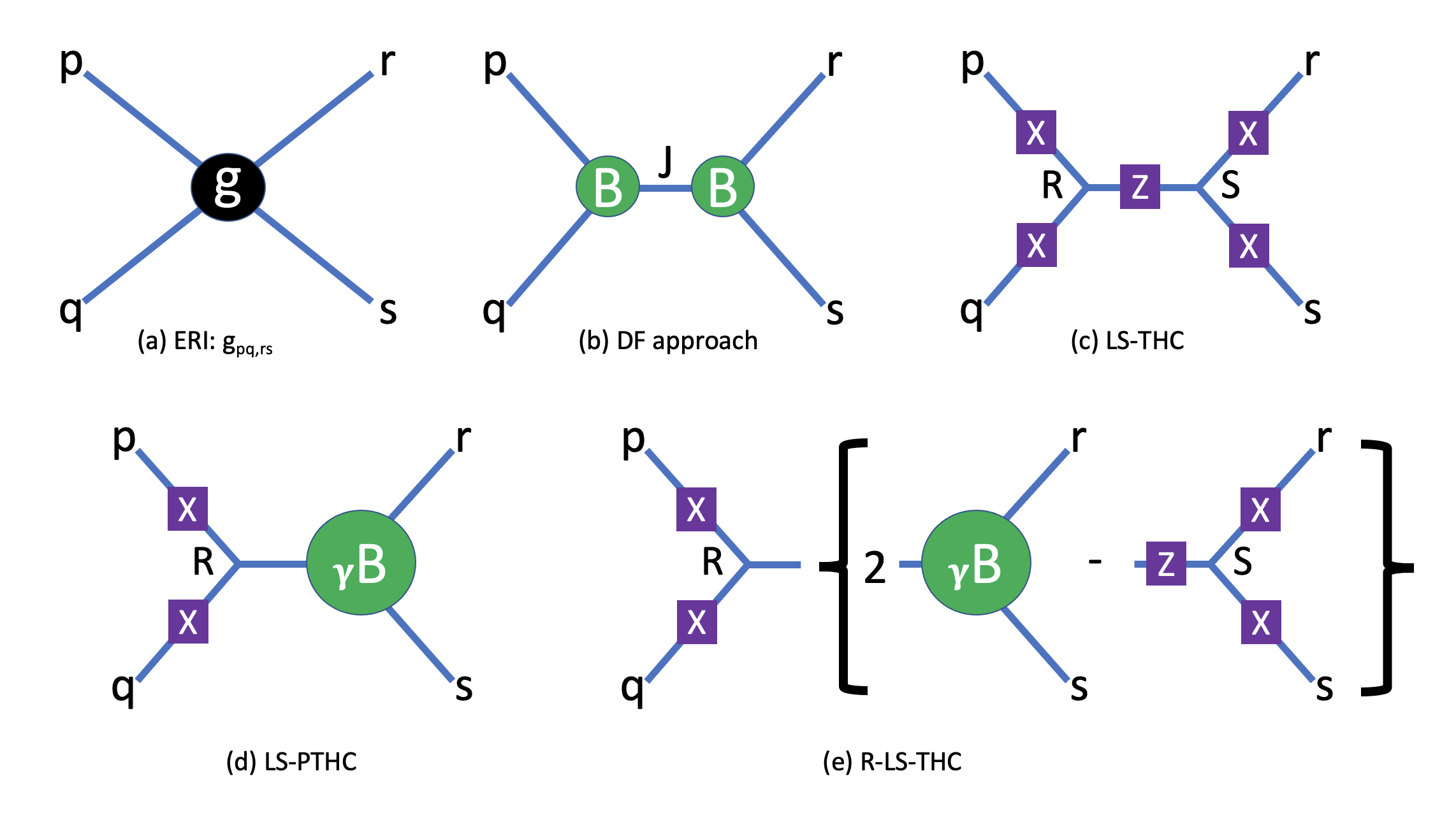}\caption{Pictorial representation of various approximations to decompose particle-particle interaction tensor term.\label{fig:schematic}}
\end{figure}

\subsection{Least-squares Tensor Hypercontraction}

The tensor representation of ERI is expressed as, 
\begin{align}
\label{ERI}
    g^{pq}_{rs} \equiv (pr|qs) = \int \phi^{*}_{p}(\mathbf{r}_1)\phi_{r}(\mathbf{r}_1) g(\mathbf{r}_1,\mathbf{r}_2)\phi^{*}_{q}(\mathbf{r}_2)\phi_{s}(\mathbf{r}_2)d\mathbf{r}_1d\mathbf{r}_2
\end{align}
here, the operator $g(\mathbf{r}_1,\mathbf{r}_2)$ = $|\mathbf{r}_1 - \mathbf{r}_2|^{-1}$ denotes the Coulomb kernel. In the least-squares THC (LS-THC) method,\cite{THC-2012,THC2-2012,THC3-2012} the ERI tensor $\mathbf{g}$ is factorized over the phyiscal grid points into five matrices (as depicted in Figure~\ref{fig:schematic}c),
\begin{align}
\label{ERI-to-THC}
    (pq|rs) \approx (pq|rs)_{THC} = \sum_{RS} X^{R}_{p}X^{R}_{q}V_{RS}X^{S}_{r}X^{S}_{s}
\end{align}
The collocation matrices \textbf{X} are determined \textit{a priori} by the evaluation of spatial MOs at grid points, $X^R_p = \phi_p(\mathbf{r}_R)$. The core matrix \textbf{V} captures the Coulomb-repulsion interaction. This matrix is determined from least squares fitting of the exact ERI tensor or some intermediate approximated tensor. Specifically as used in this work, the core matrix can be determined starting from the DF approximation with $\mathcal{O}(N^4)$ scaling where $N=N_v+N_o$,\cite{THC2-2012}
\begin{align}
    (pq|rs)_{DF} &= \sum_{KL} (pq|K) (K|L)^{-1} (L|rs)  \nonumber \\
    &= \sum_{JKL} (pq|K)(K|J)^{-1/2} (J|L)^{-1/2} (L|rs)  \nonumber\\
     \label{eq:ERI-to-DF}
    &= \sum_{J} B^{J}_{pq}B^{J}_{rs} \\
    \label{eq:core-matrix}
    V_{RS} &= \min_{V_{RS}} \frac{1}{2} \sum_{pqrs} |(pq|rs)_{DF} - X^{R}_{p}X^{R}_{q}V_{RS}X^{S}_{r}X^{S}_{s} |^2  \nonumber \\
           &= \sum_{R'S'} (S^{-1})_{R'R} E_{R'S'} (S^{-1})_{S'S} \\
    E_{RS} &= \sum_{pqrs} X^{R}_{p}X^{R}_{q} (pq|rs)_{DF} X^{S}_{r}X^{S}_{s} \nonumber \\
       &= \sum_J \left( \sum_{pq} X^{R}_{p}X^{R}_{q} B^{J}_{pq} \right) \left( \sum_{rs} X^{S}_{r}X^{S}_{s} B^{J}_{rs} \right) \nonumber \\
       &= \sum_J \eta^{RJ} \eta^{SJ} \label{eq:fitting-matrix} \\
    S_{RS} &=  \sum_{pq} X^{R}_{p}X^{R}_{q} X^{S}_{p}X^{S}_{q} 
\end{align}
The induced factorization of the fitting matrix $\mathbf{E}=\boldsymbol{\upeta}\boldsymbol{\upeta}^{\mathrm{T}}$ when fitting the DF ERI tensor also leads to a factorized form of the core matrix as well as a direct approximation of the density fitting factors,
\begin{align}
   V_{RS} &= \sum_J \left( \sum_{R'} (S^{-1})_{R'R} \eta^{R'J} \right) \left( \sum_{S'}  (S^{-1})_{S'S} \eta^{S'J} \right) \nonumber \\
   &= \sum_J \gamma^{RJ} \gamma^{SJ} \label{eq:factorized-V} \\
   B^J_{pq} &\approx \sum_R X_p^R X_q^R \gamma^{RJ} \label{eq:df-3way}
\end{align}
Note that in practice, separate core matrices are determined for different classes of ERIs $(ab|cd)$, $(ab|ci)$, $(ab|ij)$, $(ai|bj)$, $(ai|jk)$, and $(ij|kl)$ (constructed from $\boldsymbol{\upgamma}$ matrices fit to distinct classes of DF factors $B_{ab}^J$, $B_{ai}^J$, and $B_{ij}^J$), and different effective grids and hence collocation matrices are used for virtual-virtual, virtual-occupied, and occupied-occupied MO pairs.\cite{Matthews-grid-prune,Matthews-THC-MP3}

\subsection{Robust Tensor Hypercontraction}

Pierce et al. discussed various THC and THC-like approximations to the PPL term,\cite{THC-PPL-2021} which are schematically described in Figure~\ref{fig:schematic}. The central factorization used in their work is a 3-way CP factorization of the virtual-virtual density fitting factors, determined via an iterative non-linear procedure such as alternating least squares,\cite{KoldaBader}
\begin{align}
    B^{J}_{ab} \approx (B_{CP})^{J}_{ab} = \sum_{R} \beta^{R}_a \beta^{R}_{b} \gamma^{RJ}
\end{align}
Following \eqref{eq:ERI-to-DF}, this CP decomposition can be applied to one of the density fitting factor tensors leading to a ``CP-PS" factorization as illustrated in Figure~\ref{fig:schematic}d,
\begin{align}
(ab|cd)_{DF} \approx (ab|cd)_{CP-PS} &= \sum_{J} \left( \sum_{R} \beta^{R}_a \beta^{R}_{b}  \gamma^{RJ} \right) B^{J}_{cd} \nonumber \\ 
&= \sum_{R} \beta^{R}_a \beta^{R}_{b} \left( \sum_{J} \gamma^{RJ} B^{J}_{cd} \right) \nonumber \\ 
&= \sum_{R} \beta^{R}_a \beta^{R}_{b} (\gamma B)^{R}_{cd} \label{eq:CP-PS}
\end{align}
The CP-PS factorization has the same form as the partial LS-THC method (LS-PTHC)\cite{THC-PPL-2014} except that the latter is obtained by substituting $\mathbf{X}$ for $\boldsymbol{\upbeta}$ and the linear least-squares solve in \eqref{eq:factorized-V} and \eqref{eq:df-3way}. Substituting the CP decomposition for both density fitting factor tensors results in a ``CP-DF" method (Figure~\ref{fig:schematic}c, equivalent in form to LS-THC),
\begin{align}
(ab|cd)_{DF} \approx (ab|cd)_{CP-DF} &= \sum_J \left( \sum_{R} \beta^{R}_a \beta^{R}_{b} \gamma^{RJ} \right) \left( \sum_S \beta^{S}_{c} \beta^{S}_{d} \gamma^{SJ} \right) \nonumber \\
&= \sum_{RS} \beta^{R}_a \beta^{R}_{b} \left( \sum_J \gamma^{RJ} \gamma^{SJ} \right) \beta^{S}_{c} \beta^{S}_{d} \nonumber \\
&= \sum_{RS} \beta^{R}_a \beta^{R}_{b} V_{RS} \beta^{S}_{c} \beta^{S}_{d} \label{CP-DF}
\end{align}

Pierce et al. analyzed the error in each approximation by considering the errors in each component tensor (here, only the density fitting factors),
\begin{align}
    \mathbf{B} - \mathbf{B}_{CP} &= \Delta\mathbf{B} \\
    (ab|cd)_{DF} - (ab|cd)_{CP-PS} &= \sum_J (\Delta B)^J_{ab} B^J_{cd} \\
    (ab|cd)_{DF} - (ab|cd)_{CP-DF} &= \sum_J (\Delta B)^J_{ab} B^J_{cd} + \sum_J B^J_{ab} (\Delta B)^J_{cd} + \sum_J (\Delta B)^J_{ab} (\Delta B)^J_{cd}
\end{align}
Because $(ab|cd) = (cd|ab)$ the leading-order error can be canceled, leading to a robust fitting method (Figure~\ref{fig:schematic}e),
\begin{align}
    (ab|cd)_{DF} \approx (ab|cd)_{rCP-DF} &= 2 (ab|cd)_{CP-PS} - (ab|cd)_{CP-DF} \nonumber \\
&= \sum_{RS} \beta^{R}_a \beta^{R}_{b} \left( 2 (\gamma B)^{J}_{cd} - \sum_S V_{RS} \beta^S_c \beta^S_d \right) \nonumber \\
&= \sum_{RS} \beta^{R}_a \beta^{R}_{b} (\widetilde{\gamma B})^R_{cd} \label{eq:rCP-DF} \\
    (ab|cd)_{DF} - (ab|cd)_{rCP-DF} &= - \sum_J (\Delta B)^J_{ab} (\Delta B)^J_{cd}
\end{align}
assuming the approximate integrals are effectively symmetrized. It should be noted that this factorization form breaks the exact symmetry of the ERIs, although when used in the PPL, the symmetry of the double excitation amplitudes combined with explicit symmetrization of the residual compensate for this symmetry breaking (vide infra).

The definition of a robust R-LS-THC approximation follows clearly from the relationship between CP-PS/LS-PTHC and CP-DF/LS-THC by substituting $\mathbf{X}$  for $\boldsymbol{\upbeta}$ and solving for $\boldsymbol{\upgamma}$ and $\mathbf{V}$ using \eqref{eq:factorized-V} and \eqref{eq:df-3way} in the definition of \eqref{eq:rCP-DF}, rather than a non-linear CP factorization.

\subsection{The Particle-Particle Ladder Term in CCSD and EOM-CCSD}

Solution of the CCSD ground state problem revolves around the computation of the coupled cluster singles and doubles residual vectors,
\begin{align}
    Z^a_i &= \langle \Phi^a_i | \bar{H} | \Phi_0 \rangle \label{eq:z1} \\
    Z^{ab}_{ij} &= \langle \Phi^{ab}_{ij} | \bar{H} | \Phi_0 \rangle \label{eq:z2} \\
    \bar{H} &= e^{-\hat{T}} \hat{H} e^{\hat{T}} = \left( \hat{H} e^{\hat{T}} \right)_c \\
    \hat{H} &= \sum_{pq} f^p_q \hat{a}^\dagger_p \hat{a}_q + \frac{1}{2} \sum_{pqrs} g^{pq}_{rs} \hat{a}^\dagger_p \hat{a}^\dagger_q \hat{a}_s \hat{a}_r \\
    \hat{T} &= \sum_{ai} t^a_i \hat{a}^\dagger_a \hat{a}_i + \frac{1}{4} \sum_{abij} t^{ab}_{ij} \hat{a}^\dagger_a \hat{a}^\dagger_b \hat{a}_j \hat{a}_i
\end{align}
where $|\Phi_0\rangle$ is the reference determinant, $|\Phi^{ab\ldots}_{ij\ldots}\rangle = \hat{a}^\dagger_a \hat{a}^\dagger_b \ldots \hat{a}_j \hat{a}_i |\Phi_0\rangle$ are excited determinants, $\hat{a}_p$ ($\hat{a}^\dagger_p$) are MO annihilation (creation) operators, and $(\ldots)_c$ denotes a connected expression.

The CCSD singles and doubles amplitudes $t^a_i$ and $t^{ab}_{ij}$ are determined by solving for $\mathbf{Z}=0$, which in practice is realized by repeated calculation of \eqref{eq:z1} and \eqref{eq:z2} followed by adjustment of $t^a_i$ and $t^{ab}_{ij}$. Calculation of \eqref{eq:z2} is the leading-order computational cost, which in turn is commonly dominated by the PPL. In a closed-shell formalism, the doubles residual can be expressed as,
\begin{align}
    Z^{ab}_{ij} &= (1+P^{ai}_{bj}) \left( \frac{1}{2} g^{ab}_{ij} + \sum_e t^e_i g^{ab}_{ej} - \sum_m t^a_m \tilde{W}^{mb}_{ij} + \sum_{em} t^e_m (2g^{mb}_{ej} - g^{mb}_{je}) \right. \nonumber \\
    &+ \left. \sum_e F^a_e t^{eb}_{ij} - \sum_m F^m_i t^{ab}_{mj} + \frac{1}{2} \sum_{em} (2t^{ae}_{im} - t^{ae}_{mi}) (2\tilde{W}^{mb}_{ej} - \tilde{W}^{mb}_{je}) \right. \nonumber \\
    &\left. - (\frac{1}{2} + P^i_j) \sum_{em} t^{ae}_{mi} \tilde{W}^{mb}_{je} + \frac{1}{2} \sum_{mn} t^{ab}_{mn} W^{mn}_{ij} + \frac{1}{2} \sum_{ef} t^{ef}_{ij} g^{ab}_{ef} \right) \label{eq:z2rhf} \\
    F^a_e &= f^a_e + \sum_{fm} t^f_m (2g^{am}_{ef} - g^{am}_{fe}) - \sum_{fmn} t^{af}_{mn} (2g^{mn}_{ef} - g^{mn}_{fe}) - \sum_m t^a_m F^m_e \\
    F^m_i &= f^m_i + \sum_{en} t^e_n (2g^{mn}_{ie} - g^{mn}_{ei}) + \sum_{efn} t^{ef}_{in} (2g^{mn}_{ef} - g^{mn}_{fe}) + \sum_e t^e_i F^m_e \\
    F^m_e &= f^m_e + \sum_{fn} t^f_n (2g^{mn}_{ef} - g^{mn}_{fe}) \\
    \tilde{W}^{mb}_{ij} &= g^{mb}_{ij} + \sum_e t^e_i g^{mb}_{ej} + \sum_e t^e_j g^{mb}_{ie} + \sum_{ef} (t^{ef}_{ij} + t^e_i t^f_j) g^{mb}_{ef} \\
    \tilde{W}^{mb}_{ej} &= g^{mb}_{ej} - \sum_n t^b_n W^{mn}_{ej} + \sum_f t^f_j g^{mb}_{ef} + \frac{1}{4} \sum_{fn} (2t^{fb}_{nj} - t^{fb}_{jn}) (2g^{mn}_{ef} - g^{mn}_{fe}) \nonumber \\
    &- \frac{1}{4} \sum_{fn} t^{fb}_{jn} g^{mn}_{fe} \\
    \tilde{W}^{mb}_{je} &= g^{mb}_{je} - \sum_n t^b_n W^{mn}_{je} + \sum_f t^f_j g^{mb}_{fe} - \frac{1}{2} \sum_{fn} t^{fb}_{jn} g^{mn}_{fe} \\
    W^{mn}_{ej} &= g^{mn}_{ej} + \sum_f t^f_j g^{mn}_{ef} \\
    W^{mn}_{ij} &= (1 + P^{mi}_{nj}) \left( \frac{1}{2} g^{mn}_{ij} + \sum_e t^e_i g^{mn}_{ej} + \frac{1}{2} \sum_{ef} (t^{ef}_{ij} + t^e_i t^f_j) g^{mn}_{ef} \right)
\end{align}
where the permutation operator $P$ exchanges corresponding upper and lower indices in the following expression. The PPL is the last term in \eqref{eq:z2rhf}, and due to the presence of four virtual and two occupied MO indices, scales as $\mathcal{O}(N_v^4 N_o^2)$. Insertion of the R-LS-THC form of \eqref{eq:rCP-DF} into the PPL gives,
\begin{align}
    Z^{ab}_{ij} &\gets (1+P^{ai}_{bj}) \left( \frac{1}{2} \sum_{ef} t^{ef}_{ij} \sum_R X^R_a X^R_e (\widetilde{\gamma B})^R_{bf} \right) \nonumber \\
    &= (1+P^{ai}_{bj}) \left( \frac{1}{2} \sum_{R} X^R_a \left( \sum_{f} (\widetilde{\gamma B})^R_{bf} \left( \sum_e X_e^R t^{ef}_{ij} \right) \right) \right) \label{eq:ppl-factorized}
\end{align}
Inspection of the final parenthesized expression shows that this form scales as $\mathcal{O}(N_R N_v^2 N_o^2)$ where $N_R$ is the number of grid points which itself is chosen to scale linearly with the size of the system. This, the overall scaling of this term is reduced from $\mathcal{O}(N^6)$ to $\mathcal{O}(N^5)$. Note that because $t^{ef}_{ij} = t^{fe}_{ji}$ and $ef$ are dummy indices,
\begin{align}
    (1+P^{ai}_{bj}) & \left( \sum_{ef} t^{ef}_{ij} \sum_R X^R_a X^R_e (\widetilde{\gamma B})^R_{bf} \right) \nonumber \\
    &= \sum_{ef} t^{ef}_{ij} \sum_R X^R_a X^R_e (\widetilde{\gamma B})^R_{bf} + \sum_{ef} t^{ef}_{ji} \sum_R X^R_b X^R_e (\widetilde{\gamma B})^R_{af} \nonumber \\
    &= \sum_{ef} t^{ef}_{ij} \left( \sum_R X^R_a X^R_e (\widetilde{\gamma B})^R_{bf} + \sum_R X^R_b X^R_f (\widetilde{\gamma B})^R_{ae} \right)
\end{align}
and so $(ab|cd)_{R-LS-THC}$ is effectively symmetrized. When using the ``two-sided" LS-THC factorization without robust fitting, a slightly different form of the PPL may be used,
\begin{align}
    Z^{ab}_{ij} &\gets (1+P^{ai}_{bj}) \left( \frac{1}{2} \sum_{ef} t^{ef}_{ij} \sum_{RS} X^R_a X^R_e V_{RS} X^S_b X^S_f \right) \nonumber \\
    &= (1+P^{ai}_{bj}) \left( \frac{1}{2} \sum_S X^S_b \left( \sum_{R} X^R_a V_{RS} \left( \sum_{f} X^S_f \left( \sum_e X_e^R t^{ef}_{ij} \right) \right) \right) \right)
\end{align}
Even though this form performs more floating point operations ($\mathcal{O}(N_R^2 N_v N_o^2)$ compared to $\mathcal{O}(N_R N_v^2 N_o^2)$ with typically $N_R > N_v$), the required I/O from main memory is lower, and in our experiments there is a slight performance advantage to this latter form.

For an arbitrary $\lambda$-th excited state, the EOMEE-CCSD method involves determining a set of excitation amplitudes which are in fact an eigenvector of the CCSD transformed Hamiltonian with eigenvalue equal to the excited state energy $E_\lambda$,
\begin{align}
    \langle \Phi^a_i | \bar{H} \hat{R}(\lambda) | \Phi_0 \rangle &= E_\lambda \langle \Phi^a_i | \hat{R}(\lambda) | \Phi_0 \rangle \\
    \langle \Phi^{ab}_{ij} | \bar{H} \hat{R}(\lambda) | \Phi_0 \rangle &= E_\lambda \langle \Phi^{ab}_{ij} | \hat{R}(\lambda) | \Phi_0 \rangle \\
    \hat{R}(\lambda) &= r_0(\lambda) + \sum_{ai} r^a_i(\lambda) \hat{a}^\dagger_a \hat{a}_i + \frac{1}{4} \sum_{abij} r^{ab}_{ij}(\lambda) \hat{a}^\dagger_a \hat{a}^\dagger_b \hat{a}_j \hat{a}_i
\end{align}
Because the dimension of $\bar{H}$ is very large, and only a small number of eigenvectors and eigenvalues are desired, iterative diagonalization techniques such as Davidson's method\cite{Davidson} are typically used. In such an approach, the rate-limiting step is extremely similar to that in ground state CCSD, with the key quantity being the so-called $\sigma$ vector $\sigma(\lambda)|\Phi_0\rangle = [\bar{H}, \hat{R}(\lambda)]|\Phi_0\rangle$. The commutator is included to simplify the equations and directly yield the vertical excitation energy $\omega_\lambda = E_\lambda - E_{CC}$ as the eigenvalue. For a closed-shell reference, the $\sigma$ doubles vector can be computed as (dropping $\lambda$ for conciseness),
\begin{align}
    \sigma^{ab}_{ij} &= (1+P^{ai}_{bj}) \left(\sum_e r^e_i W^{ab}_{ej} - \sum_m r^a_m W^{mb}_{ij} + \sum_e F^a_e r^{eb}_{ij} - \sum_m F^m_i r^{ab}_{mj} \right. \nonumber \\
    &\left. + \sum_e G^a_e t^{eb}_{ij} - \sum_m G^m_i t^{ab}_{mj} + \frac{1}{2} \sum_{em} (2r^{ae}_{im} - r^{ae}_{mi}) (2W^{mb}_{ej} - W^{mb}_{je}) \right. \nonumber \\
    &\left. - (\frac{1}{2} + P^i_j) \sum_{em} r^{ae}_{mi} W^{mb}_{je} + \frac{1}{2} \sum_{mn} r^{ab}_{mn} W^{mn}_{ij} + \frac{1}{2} \sum_{ef} r^{ef}_{ij} W^{ab}_{ef} \right) \label{eq:s2rhf} \\
    G^a_e &= \sum_{fm} r^f_m (2W^{am}_{ef} - W^{am}_{fe}) - \sum_{fmn} r^{af}_{mn} (2g^{mn}_{ef} - g^{mn}_{fe}) \\
    G^m_i &= \sum_{en} r^e_n (2W^{mn}_{ie} - W^{mn}_{ei}) + \sum_{efn} r^{ef}_{in} (2g^{mn}_{ef} - g^{mn}_{fe}) \\
    W^{ab}_{ef} &= (1 + P^{ae}_{bf}) \left( \frac{1}{2} g^{ab}_{ef} - \sum_m t^a_m g^{mb}_{ef} + \frac{1}{2} \sum_{mn} (t^{ab}_{mn} + t^a_m t^b_n) g^{mn}_{ef} \right) \label{eq:wabef} \\
    W^{ab}_{ej} &= g^{ab}_{ej} - \sum_m t^a_m W^{mb}_{ej} - \sum_m t^b_m W^{am}_{ej} + \sum_{mn} (t^{ab}_{mn} - \frac{1}{2} t^a_m t^b_n) W^{mn}_{ej} - \sum_m F^m_e t^{ab}_{mj} \nonumber \\
    &+ \frac{1}{2} \sum_{fm} (2t^{fb}_{mj} - t^{fb}_{jm}) (2g^{am}_{ef} - g^{am}_{fe}) - (\frac{1}{2} + P^a_b) \sum_{fm} t^{fb}_{jm} g^{am}_{fe} + \sum_f t^f_j g^{ab}_{ef}  \label{eq:wabej} \\
    W^{am}_{ef} &= g^{am}_{ef} - \sum_n t^a_n g^{nm}_{ef} \\
    W^{mb}_{ej} &= 2 \tilde{W}^{mb}_{ej} - g^{mb}_{ej} + \sum_n t^b_n W^{mn}_{ej} - \sum_f t^f_j g^{mb}_{ef} \\
    W^{mb}_{je} &= 2 \tilde{W}^{mb}_{je} - g^{mb}_{je} + \sum_n t^b_n W^{mn}_{je} - \sum_f t^f_j g^{mb}_{fe} \\
    W^{mb}_{ij} &= \tilde{W}^{mb}_{ij} - \sum_n t^b_n W^{mn}_{ij} + \sum_e F^m_e t^{eb}_{ij} + \frac{1}{2} \sum_{en} (2t^{eb}_{nj} - t^{eb}_{jn}) (2W^{mn}_{ie} - W^{mn}_{ei}) \nonumber \\
    &- (\frac{1}{2} + P^i_j) \sum_{en} t^{eb}_{jn} W^{mn}_{ei}
\end{align}
Here, $(ab|cd)$ appears in two places: in the final PPL term of \eqref{eq:s2rhf} via \eqref{eq:wabef}, and also in the intermediate \eqref{eq:wabej}. While \eqref{eq:s2rhf} represents the most common and direct way to calculate the $\sigma$ doubles vector, we use a modified form which isolates the term in \eqref{eq:wabef} involving $(ab|cd)$ and allows for the elimination of all terms scaling as $\mathcal{O}(N_v^4 N_o^2)$ after factorization,
\begin{align}
    \sigma^{ab}_{ij} &= (1+P^{ai}_{bj}) \left(\sum_e r^e_i W^{ab}_{ej} - \sum_m r^a_m W^{mb}_{ij} - \sum_m t^a_m G^{mb}_{ij} + \sum_{em} r^e_m (2W^{mb}_{ej} - W^{mb}_{je})\right. \nonumber \\
    &\left. + \sum_e F^a_e r^{eb}_{ij} - \sum_m F^m_i r^{ab}_{mj} + \sum_e G^a_e t^{eb}_{ij} - \sum_m G^m_i t^{ab}_{mj} \right. \nonumber \\
    &\left. + \frac{1}{2} \sum_{em} (2r^{ae}_{im} - r^{ae}_{mi}) (2W^{mb}_{ej} - W^{mb}_{je}) - (\frac{1}{2} + P^i_j) \sum_{em} r^{ae}_{mi} W^{mb}_{je} \right. \nonumber \\
    &\left. + \frac{1}{2} \sum_{mn} r^{ab}_{mn} W^{mn}_{ij} + \frac{1}{2} \sum_{mn} (t^{ab}_{mn} + t^a_m t^b_n) G^{mn}_{ij} + \frac{1}{2} \sum_{ef} r^{ef}_{ij} g^{ab}_{ef} \right) \label{eq:s2rhf-mod} \\
    G^{mb}_{ij} &= \sum_{ef} r^{ef}_{ij} g^{mb}_{ef} \\
    G^{mn}_{ij} &= \sum_{ef} r^{ef}_{ij} g^{mn}_{ef}
\end{align}
When using the DF approximation, the presence of $(ab|cd)$ in \eqref{eq:wabej} does not present a scaling obstacle. While this term could be combined with that from \eqref{eq:wabef} by modifying the $\hat{R}_2$ amplitudes in \eqref{eq:s2rhf-mod}, $r^{ef}_{ij} \rightarrow (r^{ef}_{ij} + \tfrac{1}{2}t^e_i r^f_j + \tfrac{1}{2} r^e_i t^f_j)$, we simply compute this term by taking advantage of the DF factorization,
\begin{align}
    \sigma^{ab}_{ij} &\gets (1+P^{ai}_{bj}) \sum_{ef} r^e_i t^f_j \left( \sum_J B^J_{ae} B^J_{bf} \right) \nonumber \\
    &= (1+P^{ai}_{bj}) \sum_J \left( \sum_{e} r^e_i B^J_{ae} \right) \left( \sum_f t^f_j B^J_{bf} \right)
\end{align}
which scales as $\mathcal{O}(N_{DF} N_v^2 N_o^2)$ where $N_{DF}$ is the number of auxiliary density fitting basis functions. Now, the final PPL term of \eqref{eq:s2rhf-mod} can be computed exactly as in \eqref{eq:ppl-factorized} with the same benefits in scaling reduction.

Finally, the energy of the $\kappa$-th electron-attached state may be obtained using EOMEA-CCSD. Electron attachement amplitudes and the final state energy can be obtained by solving an eigenvalue problem as in EOMEE-CCSD, except that the eigenvector now represents a non-number-conserving operator,
\begin{align}
    \langle \Phi^a | \bar{H} \hat{R}(\kappa) | \Phi_0 \rangle &= E_\kappa \langle \Phi^a | \hat{R}(\kappa) | \Phi_0 \rangle \\
    \langle \Phi^{ab}_{i} | \bar{H} \hat{R}(\kappa) | \Phi_0 \rangle &= E_\lambda \langle \Phi^{ab}_{i} | \hat{R}(\kappa) | \Phi_0 \rangle \\
    \hat{R}(\kappa) &= \sum_{a} r^a(\kappa) \hat{a}^\dagger_a + \frac{1}{2} \sum_{abi} r^{ab}_{i}(\kappa) \hat{a}^\dagger_a \hat{a}^\dagger_b \hat{a}_i
\end{align}
For a closed-shell reference, and again modifying the working equations to fully expose the $(ab|cd)$ integrals, we arrive at our EOMEA-CCSD $\sigma$ doubles vector equations,
\begin{align}
    \sigma^{ab}_{i} &= \left(\sum_e r^e W^{ab}_{ie} - \sum_m t^a_m G^{mb}_{i} - \sum_m t^b_m G^{am}_{i} + \sum_e F^a_e r^{eb}_{i} + \sum_e F^b_e r^{ae}_{i} \right. \nonumber \\
    &\left. - \sum_m F^m_i r^{ab}_{m} - \sum_m G^m t^{ab}_{im} + \frac{1}{2} \sum_{em} (2r^{eb}_{m} - r^{be}_{m}) (2W^{ma}_{ei} - W^{ma}_{ie}) \right. \nonumber \\
    &\left. - (\frac{1}{2} + P^a_b) \sum_{em} r^{be}_{m} W^{ma}_{ie} \sum_{mn} (t^{ab}_{mn} + t^a_m t^b_n) G^{mn}_{i} + \frac{1}{2} \sum_{ef} r^{ef}_{i} g^{ab}_{ef} \right) \label{eq:s2rhf-ea} \\
    G^m &= \sum_{efn} r^{fe}_{n} (2g^{mn}_{ef} - g^{mn}_{fe}) \\
    G^{mb}_{i} &= \sum_{ef} r^{ef}_{i} g^{mb}_{ef} \\
    G^{am}_{i} &= \sum_{ef} r^{ef}_{i} g^{am}_{ef} \\
    G^{mn}_{i} &= \sum_{ef} r^{ef}_{i} g^{mn}_{ef}
\end{align}
The computation of the EOMEA-CCSD PPL term using R-LS-THC is similar to the CCSD and EOMEE-CCSD case,
\begin{align}
    \sigma^{ab}_{i} &\gets \sum_{ef} r^{ef}_{i} \sum_R X^R_a X^R_e (\widetilde{\gamma B})^R_{bf} \nonumber \\
    &= \sum_{R} X^R_a \left( \sum_{f} (\widetilde{\gamma B})^R_{bf} \left( \sum_e X_e^R r^{ef}_{i} \right) \right) \label{eq:ppl-ea-factorized}
\end{align}
Here, the scaling is reduced from $\mathcal{O}(N_v^4 N_o)$ to $\mathcal{O}(N_R N_v^2 N_o)$.

Because we start with the density fitting factors $B_{ab}^J$ for fitting the $\boldsymbol{\upgamma}$ and $\mathbf{V}$ matrices, we also utilize density fitting in the remainder of the CCSD and EOM-CCSD equations. This means that with a large enough grid (such that the error due to THC fitting is negligible), the error introduced relative to canonical (EOM-)CCSD is purely due to density fitting and the same as for other available DF-(EOM-)CCSD implementations. While introducing density fitting for the complete (EOM-)CCSD equations does modify the CCSD residual and EOM-CCSD $\sigma$ equations from \eqref{eq:z2rhf}, \eqref{eq:s2rhf-mod}, and \eqref{eq:s2rhf-ea}, the details are beyond the scope of this work and do not affect the formal leading-order scaling of the remaining terms. Because all THC methods used in this work are based on the DF approximation, we drop any explicit ``DF-" modifier in the discussion and assume that density fitting is always used unless otherwise indicated.

\section{Computational Details}

DF-EOM-CCSD and the DF-based LS-THC, LS-PTHC, and R-LS-THC approximations to the PPL term were implemented in a development version of CFOUR.\cite{Matthews2020} We denote the THC-based methods as LS-THC-PPL-EOM-CCSD, LS-PTHC-PPL-EOM-CCSD, and R-LS-THC-PPL-EOM-CCSD, or simply THC-PPL-EOM-CCSD to generically refer to any of these approximations. All calculations were performed with Dunning's correlation-consistent polarized valence triple-$\zeta$ basis set with augmented diffuse functions (aug-cc-pVTZ),\cite{Dunning-tz} except for the stacked pyrimidine nucleotide system where cc-pVDZ\cite{Dunning-dz} was employed. Density fitting was performed with standard auxiliary basis sets paired to the orbital basis set (aug-cc-pVTZ-RI and cc-pVDZ-RI, respectively),\cite{auxiliary-basis} except where indicated. The SG0\cite{SG1-Gill-1993} parent grid was used for THC fitting, with a varying numerical cut-off $\epsilon$ ranging from 1.0 to 0.01 in order to control the size and accuracy of the pruned grid.\cite{Matthews-grid-prune} 

\section{Results and Discussion}

\subsection{Performance} \label{sec:timings}

\begin{figure}
\includegraphics[scale=0.6]{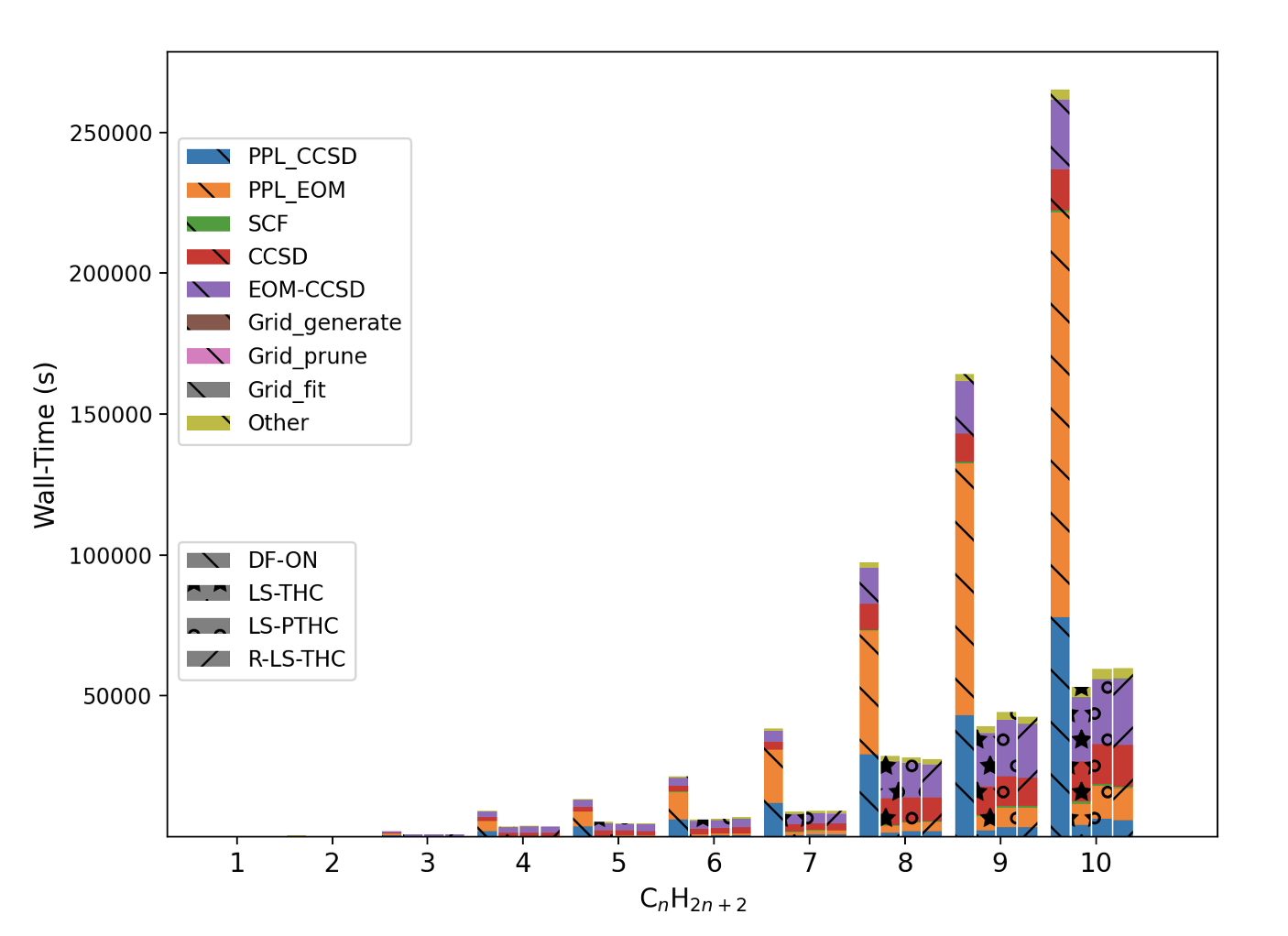}\caption{Walltimes for DF-EOMEE-CCSD (DF-ON) and various THC-PPL-EOMEE-CCSD methods for linear alkane chains. One excited state is computed in each case. The grid tolerance was set to $\epsilon = 0.01$. Timings are split into components as discussed in the text. \label{fig:performance}}
\end{figure}

A set of linear alkanes was chosen to study the performance of EOM-CCSD calculations of a single excited state. Timings are measured running on $2\times$ Intel\textsuperscript{\textregistered} Xeon\textsuperscript{\textregistered} E5-2695v4 CPUs and either 256 or 768 GiB of memory. OpenMP was used to parallelize the calculation over all 36 cores. The total time for the calculation is divided into several categories: the time taken for the self-consistent field (SCF) solution, the total time for the ground CCSD and excited state EOM-CCSD solutions less the PPL term, the time taken for evaluating the PPL term for both CCSD and EOM-CCSD parts, grid generation and pruning times, and the time required for determining the fitting parameters ($\boldsymbol{\upgamma}$, $\mathbf{V}$, $\boldsymbol{\upgamma}\mathbf{B}$, and $\widetilde{\boldsymbol{\upgamma}\mathbf{B}}$). The remaining calculation time was grouped together and indicated as ``Other". Figure~\ref{fig:performance} illustrates timings for DF-EOMEE-CCSD calculations with standard, LS-THC, LS-PTHC, and R-LS-THC PPL terms as a function of alkane chain length.

The PPL evaluation takes significant time for CCSD and EOM-CCSD calculations without THC. This time is significantly reduced for all the THC methods, by approximately a factor of 12--30$\times$. The results indicate that for the calculation of excited state, the time spend to evaluate PPL term is larger than that for the ground state; this is primarily due to the larger number of iterations taken to solve the excited state eigenvalue problem. When multiple excite states are desired this imbalance would grow even further, highlighting the need to reduce the computational cost of the EOM-CCSD stage of the calculation. Adding in the rest of the computational time, the reduction of cost for the PPL term still results in significant total walltime savings, by approximately a 5$\times$ ratio for decane. As mentioned previously, the non-robust LS-THC PPL term is slightly more computationally efficient than for LS-PTHC or R-LS-THC.

Calculations utilizing the THC approximations require additional operations related to the grid (generation, fitting, and pruning) and fitting of the initial ERIs/density fitting factors, the cost of which is not present in standard DF-EOM-CCSD calculations. Figure~\ref{fig:performance} shows that these THC-specific timings are insignificant contributions to overall walltime. This highlights one advantage of the LS-THC method in that a costly non-linear solve is not required, which can take 5--20\% of the total time in an rCP-DF calculation.\cite{THC-PPL-2021}

Finally, we note that a rather strict grid tolerance parameter $\epsilon = 0.01$ was used for these timings, while a milder tolerance of $\epsilon = 0.1$ is sufficient for $\sim$meV errors and would even further reduce the time required for the PPL. However, since even with a tight tolerance the PPL is a minor contributor to total time, further reductions in walltime will be more modest.

\subsection{Accuracy of THC-PPL-EOMEE-CCSD}

The accuracy of the THC-PPL-EOMEE-CCSD methods for vertical excitation energies was benchmarked using the QUEST data sets,\cite{quest-review} in particular, the QUEST1,\cite{quest1} QUEST3,\cite{quest3} and QUEST4\cite{quest4} data sets. We included all excited states listed in the database in our comparisons, and checked relative energies, oscillator strengths, and (where possible) assignments in terms of MOs between THC-PPL-EOMEE-CCSD, DF-EOM-CCSD, and the QUEST database to verify the selection of states.

\begin{figure}
\includegraphics[scale=0.4]{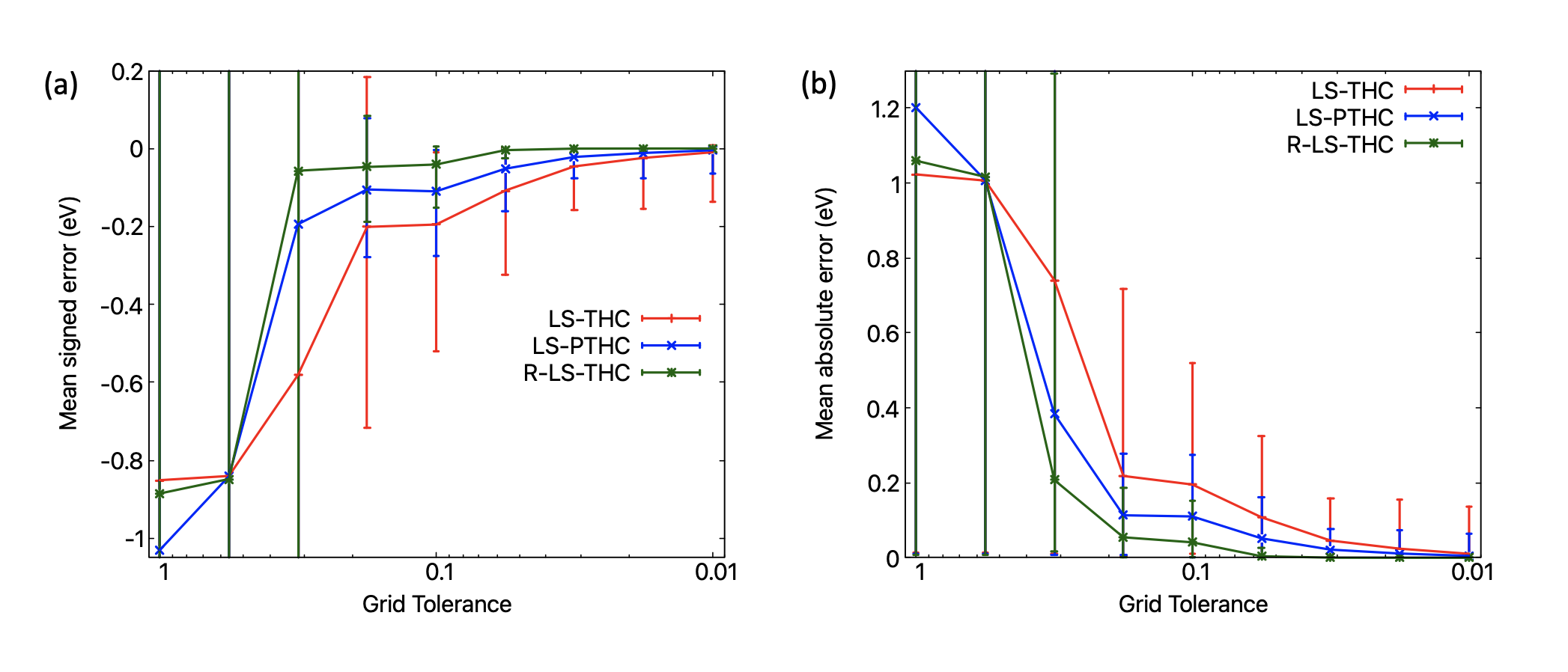}\caption{(a) Mean signed error (MSE) and (b) mean absolute error (MAE) for THC-PPL-EOMEE-CCSD vertical excitation energies calculated for excited states in the QUEST1 database as a function of grid tolerance parameter $\epsilon$. Maximum and minimum errors are indicated by ``whiskers". See text for details. \label{fig:quest1}}
\end{figure}

\begin{figure}
\includegraphics[scale=0.4]{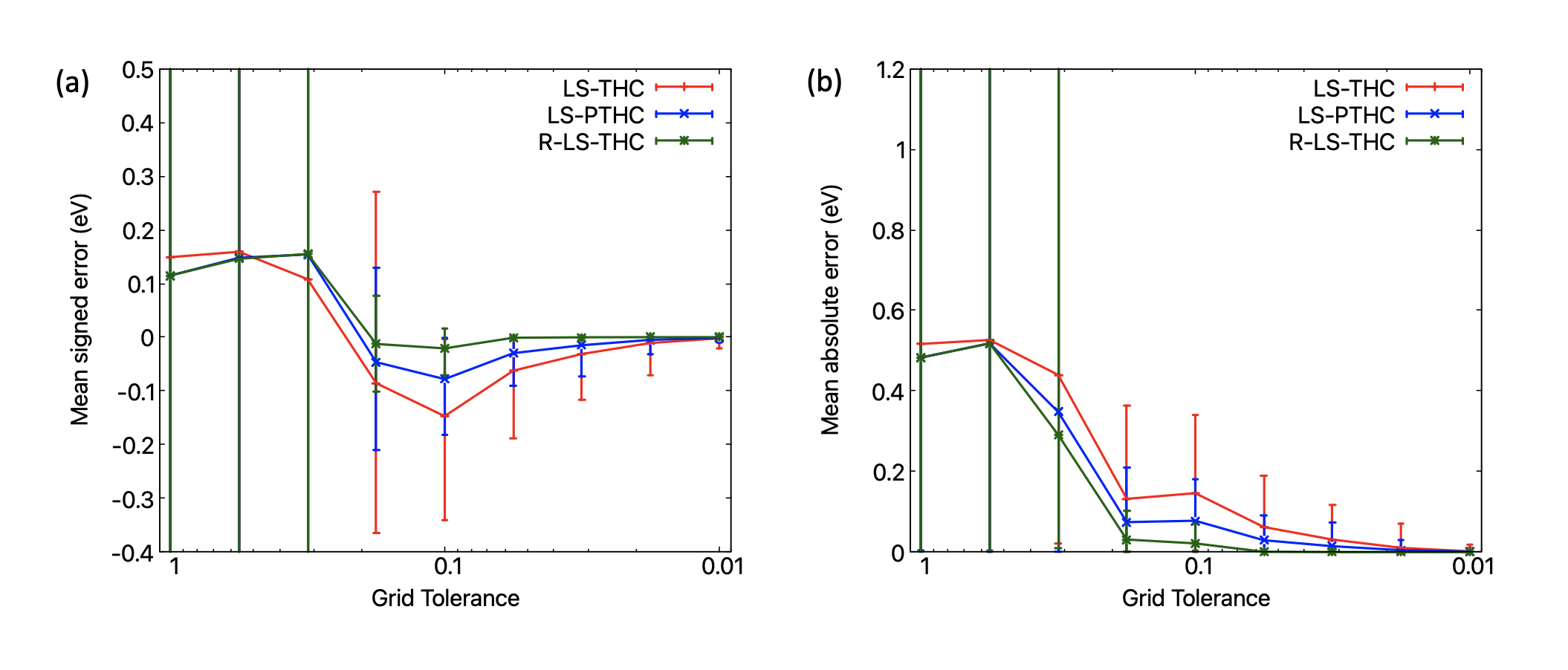}\caption{(a) Mean signed error (MSE) and (b) mean absolute error (MAE) for THC-PPL-EOMEE-CCSD vertical excitation energies calculated for excited states in the QUEST3 database as a function of grid tolerance parameter $\epsilon$. Maximum and minimum errors are indicated by ``whiskers". See text for details. \label{fig:quest3}}
\end{figure}

\begin{figure}
\includegraphics[scale=0.4]{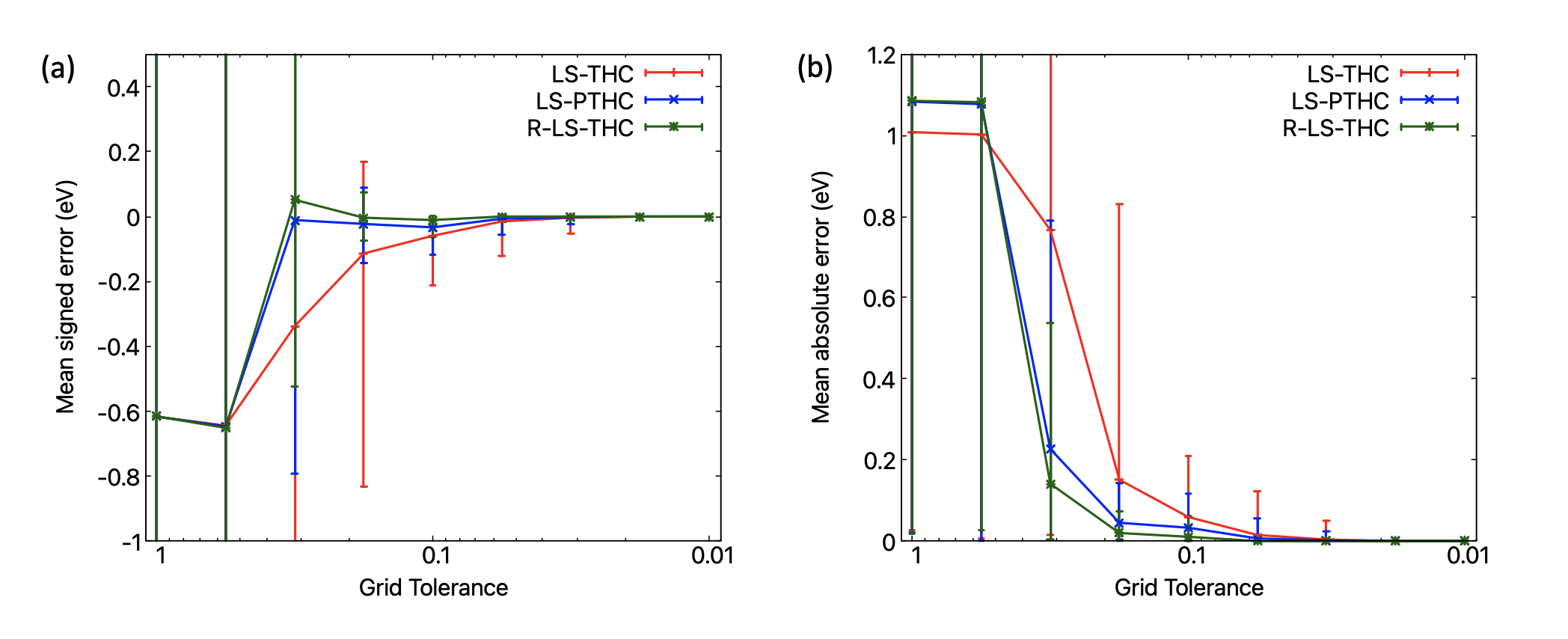}\caption{(a) Mean signed error (MSE) and (b) mean absolute error (MAE) for THC-PPL-EOMEE-CCSD vertical excitation energies calculated for excited states in the QUEST4 database as a function of grid tolerance parameter $\epsilon$. Maximum and minimum errors are indicated by ``whiskers". See text for details. \label{fig:quest4}}
\end{figure}

Reported errors are measured as differences in the excitation energy between the DF-EOMEE-CCSD and THC-PPL-EOMEE-CCSD calculations, i.e. $E_{THC-EOM-CCSD} - E_{DF-EOM-CCSD}$, for each of LS-THC, LS-PTHC, and R-LS-THC. Statistics of the errors over the QUEST sub-databases are presented in Figures~\ref{fig:quest1}, \ref{fig:quest3}, and \ref{fig:quest4}.

Figure~\ref{fig:quest1}a shows the mean signed error (MSE) for electronic excited states from the QUEST1 database. The vertical excitation energies of 18 molecules (up to 3 non-hydrogen atoms) are included to form the QUEST1 database.\cite{quest1} The average error is observed to converge for all THC approximations with tighter grid tolerance. A similar trend observed for mean absolute error (MAE, Figure~\ref{fig:quest1}b). While the errors for the R-LS-THC method reduce to $<\SI{4}{meV}$ for $\epsilon < 10^{-1}$, for the LS-PTHC and LS-THC methods the error reaches $\sim\SI{}{meV}$ accuracy only for $\epsilon = 10^{-2}$. This result points out that the error cancellation embodied in the R-LS-THC approximation can be effectively leveraged for evaluating vertical excitation energies with relatively larger grid tolerances, although as seen above, even a ``safe" value of $\epsilon = 10^{-2}$ results in large overall speedups.

The mean signed and absolute errors for excitation energies contained from the QUEST3 database are presented in Figure~\ref{fig:quest3}. The QUEST3 database consists of vertical excitation energies of molecules containing 4 to 6 non-hydrogen atoms.\cite{quest3} The average error shows an oscillatory behavior versus grid tolerance which is not present in the errors observed for QUEST1. In the mean absolute error, this change in sign of the absolute error manifests as a plateau in convergence of the error with grid size. However, average errors with very small grids ($\epsilon > 0.3$) are approximately only half as large as in QUEST1 such that the errors manifested for $0.1 \ge \epsilon \ge 0.01$ closely mirror those in QUEST1. In both cases, errors for R-LS-THC are in the single-digit meV range below $\epsilon = 0.1$.

The QUEST4 database include vertical transition energies of 30 closed shell ``exotic" molecules containing F, Cl, P, Si, etc. atoms.\cite{quest4} Figure~\ref{fig:quest4} shows the mean signed and mean absolute errors in the calculations of these excitation energies using THC approximated methods, with respect to DF-EOM-CCSD. The R-LS-THC method---as expected---exhibits a rapid convergence of the error, reaching sub-meV errors for $\epsilon < 0.1$. Overall, the observed errors follow a similar pattern as for QUEST1. However, it is interesting to note that the average errors associated with all three THC approximated methods shrink below \SI{10}{meV} as the grid tolerance passes $\epsilon = 0.1$. Maximum errors for LS-THC and LS-PTHC remain as large as 0.25 eV in this range, but the error cancellation inherent in R-LS-THC reduces maximum errors to only 63 meV (in this case, for silylidene). 

The measurement of R-LS-THC errors in the excitation energies of QUEST species indicate that for a sufficiently small grid tolerance (below $\sim 10^{-1.5}$) absolute errors in vertical excitation energies are reliably below \SI{1}{meV}. Further, it is observed that in each case the error cancellation built into the R-LS-THC method results in rapid decay of the error with decreasing grid tolerance, and significantly lower errors than for LS-THC or LS-PTHC. At low to intermediate grid tolerances ($10^{0} < \epsilon < 10^{-1.5}$), further improvement may be possible, e.g. via orbital weighting,\cite{THC-PPL-2014} or by using separate tolerances for the ground and excited state PPL terms so that errors in the less computationally-demanding ground state CCSD equations can be converged more tightly than the excited state itself (since EOMEE-CCSD relies on cancellation of missing correlation effects of the ground and excited states via the cluster amplitudes $\hat T$). However, even in the least accurate method (LS-THC-PPL-EOMEE-CCSD), simply tightening the grid tolerance parameter reliably results in negligible energy errors and, as measured at $\epsilon = 0.01$, significant reductions in walltime.

\subsection{Accuracy of THC-PPL-EOMEA-CCSD}

Similarly to electronic excitation energies, electron attachment energies (electron affinities) can be evaluated using the EOM framework.\cite{EOM} We studied the error in vertical electron attachment energies using LS-THC, LS-PTHC, and R-LS-THC approximations of the PPL, measured with respect to DF-EOMEA-CCSD. We also tested THC-PPL approximations applied to EOMEA-MBPT2; in this method the diagonalization of the transformed Hamiltonian is performed precisely as in EOMEA-CCSD, but the coupled cluster amplitudes are taken from second-order perturbation theory\cite{EOM-MBPT2} rather than a CCSD ground state calculation. Because the EOMEA stage of the calculation only scales as $\mathcal{O}(N^5)$, the replacement of CCSD with MBPT2 results in an overall scaling of $\mathcal{O}(N^5)$ rather than $\mathcal{O}(N^6)$. This also results in the EOMEA PPL term becoming the dominant computational cost.

The accuracy of these implementations was assesed by calculating the lowest vertical electron attachment energies (VEAs) for the set of 22 molecules included in the NAB22 test set.\cite{NAB-dataset} VEAs were calculated for the optimized geometries of neutral species as provided in the supplementary information of Ref.~\citenum{NAB-dataset}. The reported errors in electron attachment energies are defined in a similar manner as for excitation energies.

\begin{figure}
\includegraphics[scale=0.4]{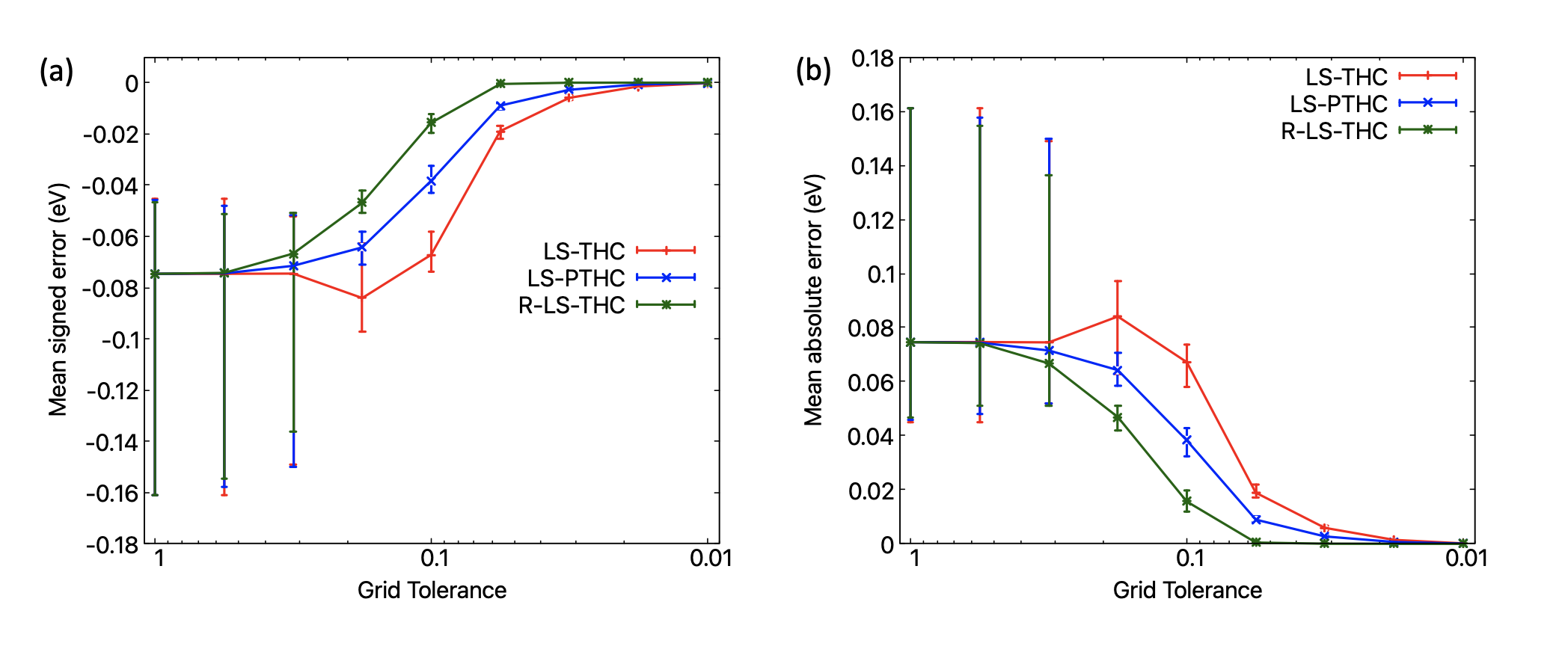}\caption{(a) Mean signed error (MSE) and (b) mean absolute error (MAE) for THC-PPL-EOMEA-CCSD vertical electron attachment energies in the NAB22 database as a function of grid tolerance parameter $\epsilon$. Maximum and minimum errors are indicated by ``whiskers". See text for details. \label{fig:EA-CCSD}}
\end{figure}

The errors in the electron attachment energies evaluated using THC-PPL-EOMEA-CCSD methods are presented in Figure~\ref{fig:EA-CCSD}. It is notable that errors for all the THC-approximated methods at all grid tolerances are negative, i.e. the THC methods consistently underestimate the EA energy in EOMEA-CCSD calculations. Thus, Figure~\ref{fig:EA-CCSD}a and Figure~\ref{fig:EA-CCSD}b show exactly same behavior. Further, it is observed that the R-LS-THC approximation leads to reduction of error below \SI{10}{meV} for $\epsilon < 10^{-1}$ and below \SI{1}{meV} for $\epsilon < 10^{-1.25}$.

\begin{figure}
\includegraphics[scale=0.4]{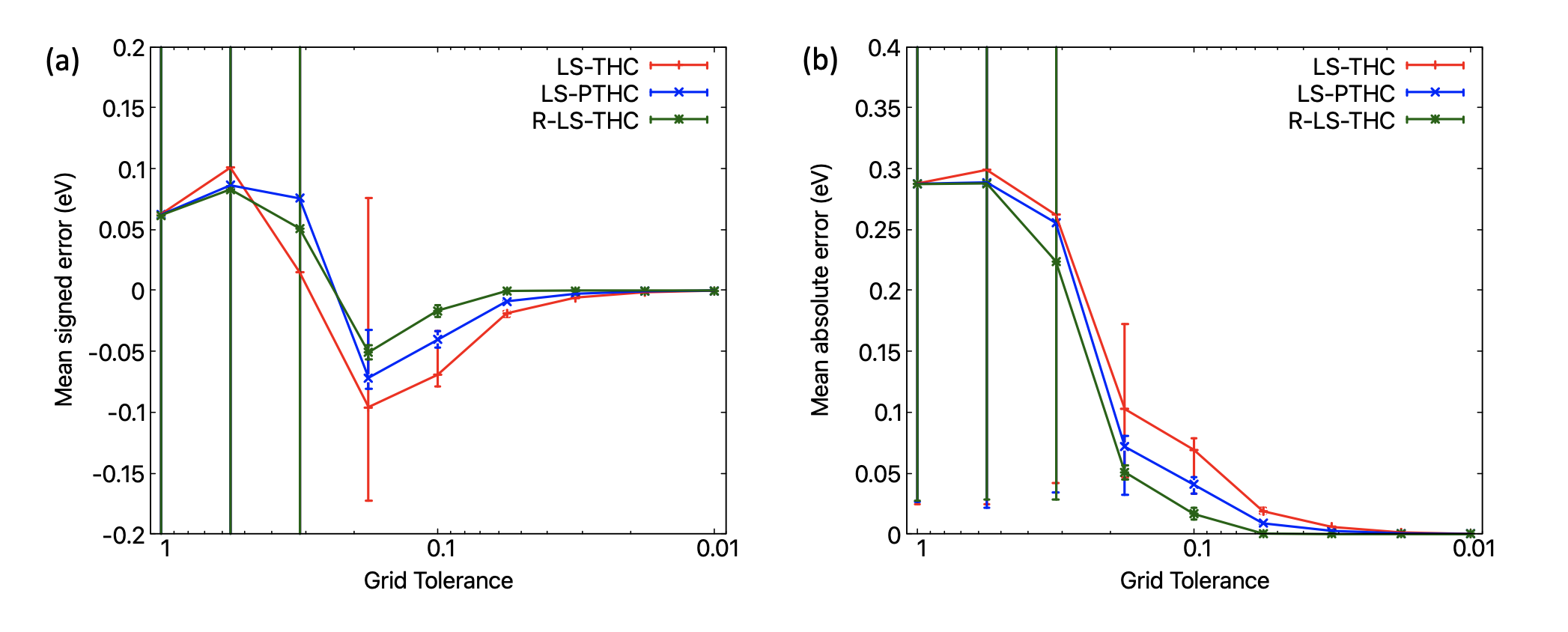}\caption{(a) Mean signed error (MSE) and (b) mean absolute error (MAE) for THC-PPL-EOMEA-MBPT2 vertical electron attachment energies in the NAB22 database as a function of grid tolerance parameter $\epsilon$. Maximum and minimum errors are indicated by ``whiskers". See text for details. \label{fig:EA-MP2}}
\end{figure}

Errors for VEAs calculated using THC-PPL-EOMEA-MBPT2 methods are presented in Figure~\ref{fig:EA-MP2}. The behavior of the errors here are seen to be quite different both qualitatively and quantitatively from those for EOMEA-CCSD. Qualitatively, the signed EOMEA-MPBT2 errors undergo a sign change near $\epsilon = 10^{-0.75}$, and exhibit a much wider range between the maximum and minimum. Mean absolute errors are also much larger for $\epsilon > 0.1$, reaching nearly $4\times$ the magnitude at $\epsilon = 1.0$. However, errors for smaller grid thresholds, below $\epsilon < 0.1$ are much more similar and converge to sub-meV values at nearly the same rate. The larger differences at loose grid thresholds may indicate a lack of balancing effect from the more accurate CCSD ground state wavefunction. Focusing only on reliable values of $0.1 > \epsilon > 0.01$, the trend in errors is essentially the same as for vertical excitation energies, but the magnitude of errors is much reduced, by as much as an order of magnitude. In part, this reflects the smaller magnitude of the electron attachment energies themselves, but there is likely also some effect from the expected simpler electronic structure of primary electron-attached wavefunctions compared to electronic excitations.

\subsection{Influence of the Density Fitting Auxiliary Basis Set}

\begin{table}
\caption{Error in the calculated excitation energy (in meV) for the lowest excited state of \ce{CH3CHO} with LS-THC-PPL- and DF-EOMEE-CCSD with respect to canonical EOMEE-CCSD. $-\log(\epsilon)=\infty$ indicates a standard DF-EOMEE-CCSD calculation, and the number of density fitting auxiliary basis functions and LS-THC grid points are indicated by $N_{DF}$ and $N_R$, respectively. Auxiliary basis sets are abbreviated, e.g. aTZ-RI = aug-cc-pVTZ-RI.}
\label{table:LS-THC-comprehensive}
\begin{tabular}{ S S S S r }
\toprule
\multicolumn{1}{c}{$-\log(\epsilon)$} & \multicolumn{1}{c}{aTZ-RI} & \multicolumn{1}{c}{aQZ-RI} & \multicolumn{1}{c}{a5Z-RI} & \multicolumn{1}{c}{$N_R$} \\
\midrule
0.00  &  -562.  &  -562.  &   -562. & 1 \\
0.25  &  -563.  &  -562.  &   -562. & 3 \\
0.50  &  -752.  &  -752.  &   -752. & 35 \\
0.75  &  -237.  &  -237.  &   -237. & 257 \\
1.00  &  -34.3   &  -34.5   &   -34.6  & 567 \\
1.25  &  -4.90    &  -5.25    &   -5.33   & 975 \\
1.50  &  -0.496    &  -0.856    &   -0.938   & 1300 \\
1.75  &   0.081    &  -0.273    &   -0.035   & 1565 \\
2.00  &  0.423     &   0.1071   &   -0.004   & 1816 \\
$\infty$    &   0.444  &   0.092    &   0.017 & --- \\
\midrule
$N_{DF}$ & \multicolumn{1}{r}{502} & \multicolumn{1}{r}{824} & \multicolumn{1}{r}{1234} & \\
\bottomrule
\end{tabular}
\end{table}

\begin{table}
\caption{Error in the calculated excitation energy (in meV) for the lowest excited state of \ce{CH3CHO} with LS-PTHC-PPL- and DF-EOMEE-CCSD with respect to canonical EOMEE-CCSD. $-\log(\epsilon)=\infty$ indicates a standard DF-EOMEE-CCSD calculation, and the number of density fitting auxiliary basis functions and LS-THC grid points are indicated by $N_{DF}$ and $N_R$, respectively. Auxiliary basis sets are abbreviated, e.g. aTZ-RI = aug-cc-pVTZ-RI.}
\label{table:LS-PTHC-comprehensive}
\begin{tabular}{ S S S S r }
\toprule
\multicolumn{1}{c}{$-\log(\epsilon)$} & \multicolumn{1}{c}{aTZ-RI} & \multicolumn{1}{c}{aQZ-RI} & \multicolumn{1}{c}{a5Z-RI} & \multicolumn{1}{c}{$N_R$} \\
\midrule
0.00  &  -563.  &  -562.  &   -562. & 1 \\
0.25  &  -565.  &  -564.  &   -564. & 3 \\
0.50  &  -503.  &  -503.  &   -503. & 35 \\
0.75  &  -121.  &  -121.  &   -121. & 257 \\
1.00  &  -15.3   &  -15.6   &   -15.6  & 567 \\
1.25  &  -2.03    &  -2.37    &   -2.44   & 975 \\
1.50  &   0.000    &  -0.353    &   -0.428   & 1300 \\
1.75  &   0.027    &  -0.081    &   -0.169   & 1565 \\
2.00  &  0.438     &   0.083   &    0.007   & 1816 \\
$\infty$    &   0.444  &   0.092    &   0.017 & --- \\
\midrule
$N_{DF}$ & \multicolumn{1}{r}{502} & \multicolumn{1}{r}{824} & \multicolumn{1}{r}{1234} & \\
\bottomrule
\end{tabular}
\end{table}

\begin{table}
\caption{Error in the calculated excitation energy (in meV) for the lowest excited state of \ce{CH3CHO} with R-LS-THC-PPL- and DF-EOMEE-CCSD with respect to canonical EOMEE-CCSD. $-\log(\epsilon)=\infty$ indicates a standard DF-EOMEE-CCSD calculation, and the number of density fitting auxiliary basis functions and LS-THC grid points are indicated by $N_{DF}$ and $N_R$, respectively. Auxiliary basis sets are abbreviated, e.g. aTZ-RI = aug-cc-pVTZ-RI.}
\label{table:R-LS-THC-comprehensive}
\begin{tabular}{ S S S S r }
\toprule
\multicolumn{1}{c}{$-\log(\epsilon)$} & \multicolumn{1}{c}{aTZ-RI} & \multicolumn{1}{c}{aQZ-RI} & \multicolumn{1}{c}{a5Z-RI} & \multicolumn{1}{c}{$N_R$} \\
\midrule
0.00  &  -563.  &  -562.  &   -562. & 1 \\
0.25  &  -566.  &  -566.  &   -566. & 3 \\
0.50  &  -351.  &  -350.  &   -350. & 35 \\
0.75  &  -43.5  &  -43.4  &   -43.3 & 257 \\
1.00  &  -0.101   &  -0.378   &   -0.382  & 567 \\
1.25  &   0.455    &  0.121    &   0.065   & 975 \\
1.50  &   0.465    &   0.119    &    0.050   & 1300 \\
1.75  &   0.440    &   0.099    &    0.026   & 1565 \\
2.00  &  0.446     &   0.095   &    0.020   & 1816 \\
$\infty$    &   0.444  &   0.092    &   0.017 & --- \\
\midrule
$N_{DF}$ & \multicolumn{1}{r}{502} & \multicolumn{1}{r}{824} & \multicolumn{1}{r}{1234} & \\
\bottomrule
\end{tabular}
\end{table}

While the accuracy of the LS-THC, LS-PTHC, or R-LS-THC approximation can be effectively controlled by the grid threshold $\epsilon$, the density fitting approximation itself also incurs an error in the excitation or electron attachment energy. Because the rate-limiting steps of the THC-PPL-EOMEE-CCSD methods do not depend strongly on the size of the auxiliary density fitting basis set (remaining terms scale as $\mathcal{O}(N_{DF} N_v^3 N_o)$ or lower, where $N_{DF}$ is the number of auxiliary basis functions), it is an intriguing possibility to increase the size of the density fitting basis set in order to further reduce error with respect to the canonical EOM-CCSD methods as well as to study the relationship between the THC and DF errors.

Tables~\ref{table:LS-THC-comprehensive}, \ref{table:LS-PTHC-comprehensive}, and \ref{table:R-LS-THC-comprehensive} show the error in calculating the lowest vertical excitation energy of acetaldehyde using LS-THC-PPL-, LS-PTHC-PPL-, and R-LS-THC-PPL-EOMEE-CCSD calculations respectively. This analysis is performed for varying $\epsilon$ as well as auxiliary basis set (while maintaining the same aug-cc-pVTZ orbital basis set). The auxiliary basis set is varied from aug-cc-pVTZ-RI to aug-cc-pV5Z-RI, to understand the convergence of DF approximation upon increasing the number of auxiliary functions. We also include standard DF-EOMEE-CCSD VEEs as the limit of an infinite THC grid size. It is observed that for DF-EOMEE-CCSD alone, the error in excitation energy converges smoothly with auxiliary basis set size, and is already rather small when using the standard aug-cc-pVTZ-RI basis set.

For the THC approximations, increasing the number of auxiliary functions does not effectively reduce the total error for larger values  of $\epsilon > 0.1$. For LS-THC, the smooth behavior of the convergence of the error with increasing auxiliary basis set is only observed for $\epsilon = 10^{-2}$, while for R-LS-THC the error seems to be smoothly convergent for $\epsilon < 10^{-1.25}$. In several cases, the THC and DF errors are of similar magnitude and opposite sign, resulting in error cancellation (e.g. $\epsilon = 10^{-1.5}$ with LS-PTHC-PPL-EOMEE-CCSD and aug-cc-pVTZ-RI). Even when of the same sign, the THC and DF errors typically combine in an uncorrelated fashion, indicating that they are largely independent. The independence of the two errors also allows us to estimate at what point they become similar in magnitude based on the relative number of auxiliary basis function and grid points. For LS-THC and LS-PTHC, the errors become similar when $N_R \sim 3N_{DF}$, while for R-LS-THC, errors are similar when $N_R \sim N_{DF}$. This relationship has interesting implications for R-LS-THC-PPL-EOMEE-CCSD with the aug-cc-pVTZ-RI auxiliary basis set and $\epsilon = 0.1$, as this is both the largest value of $\epsilon$ which produces errors on the order of \SI{1}{meV}, but is also a point of cancellation in this example. If the relative sign of the DF and THC errors is consistent across many different systems, then improving the total error by increasing the auxiliary basis set would also require decreasing $\epsilon$ (e.g. note that the total error increases almost $4\times$ when moving to aug-cc-pVQZ-RI in this case).

Overall, these results show that indeed errors can be systematically reduced by increasing the size of the density fitting auxiliary basis set and the THC grid in a concerted fashion. However, given the already extremely small error due to density fitting, there doesn't seem to be a need to go beyond standard auxiliary basis sets in most instances.

\subsection{Applicability to Large Systems}

\begin{figure}
\includegraphics[scale=0.4]{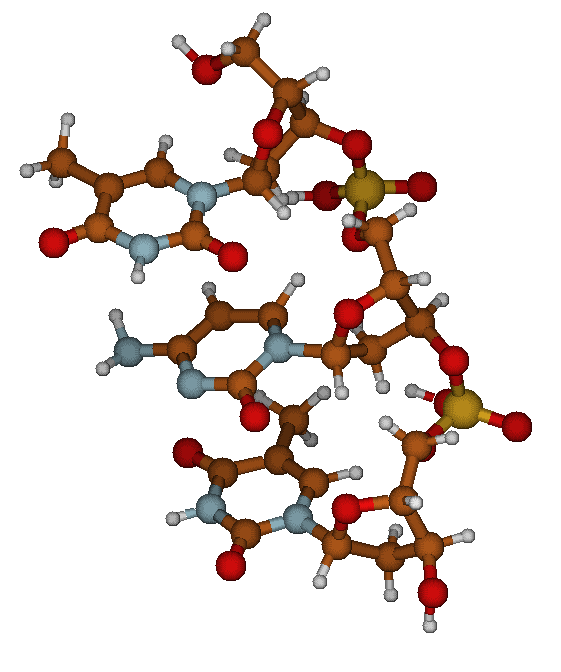}\caption{Model ssDNA system containing stacked pyrimidine nucleotides (TCT) and phosphodeoxyribose side chain. \label{fig:model}}
\end{figure}

The applicability of the THC approximated PPL term was tested for the EOMEA-MBPT2 method on single-stranded DNA (ssDNA) model system. This model system (Figure~\ref{fig:model}) includes three pyrimidine bases (thymine--cytosine--thymine: TCT) along with the phosphodeoxyribose side chain. To obtain the geometry for this model system, we started with the geometry for the DNA model system of the Ref.~\citenum{DNA-stack-geometry} and removed all atoms associated with the purine strand. The final model system consists of 95 atoms, 436 electrons, and 987 basis functions using the cc-pVDZ basis set. Three vertical electron affinities were calculated at the DF-, LS-THC-PPL-, LS-PTHC-PPL-, and R-LS-THC-PPL-EOMEA-MBPT2 level. These calculations were performed on a system with $4\times$ Intel\textsuperscript{\textregistered} Xeon\textsuperscript{\textregistered} E7-8891v3 CPUs and 1.5 TiB of memory. OpenMP was used to parallelize the calculation over all 40 cores.

\begin{table}
\caption{Primary vertical electron affinities of the ssDNA TCT model using THC-PPL and DF approximations. VEAs for DF-EOMEA-MBPT2 and differences of THC-PPL-EOMEA-MBPT2 VEAs with respect to DF-EOMEA-MBPT2 are reported in eV. The grid threshold $\epsilon$ was set to 0.01 for all THC calculations. Timing breakdowns (in hours) are also reported for various components ($t_{\mathrm{EOMEA}}$ is the total time for the EOMEA portion of the calculation less the PPL term).}
\label{table:Model-system-EA}
\begin{tabular}{ c S[table-format=2.4] S[table-format=2.2e1] S[table-format=3.2e1] S[table-format=3.2e1] }
\toprule
        & \multicolumn{1}{c}{DF} & \multicolumn{1}{c}{LS-THC} & \multicolumn{1}{c}{LS-PTHC} & \multicolumn{1}{c}{R-LS-THC} \\
\midrule
    1\textsuperscript{st} VEA & -0.719 & -3.29e-4  &  -1.64e-4  &  -1.64e-7    \\
    2\textsuperscript{nd} VEA &  0.587 & +3.86e-4  &  +1.94e-4  &  +3.85e-8    \\
    3\textsuperscript{rd} VEA &  1.176 & -1.82e-5  &  -8.79e-5  &  -5.50e-8    \\
    $t_{\mathrm{Total}}$ & 85.9 & 30.2 & 31.5 & 31.9 \\
    $t_{\mathrm{PPL}}$   & 54.9 & 0.958 & 2.88 & 2.86 \\
    $t_{\mathrm{EOMEA}}$ & 24.4 & 22.7 & 21.7 & 21.8 \\
    $t_{\mathrm{Other}}$ & 6.54 & 6.57 & 6.93 & 7.23 \\
\bottomrule
\end{tabular}
\end{table}

The calculated VEAs and timing breakdowns are presented in Table~\ref{table:Model-system-EA}. All three THC approximations feature errors below \SI{1}{meV}, with the R-LS-THC approximation reducing the error to $\sim \SI{1e-7}{eV}$ or below. The total time of the calculation is reduced by approximately a factor of 3 with each THC method. Because of the small virtual space, the remaining EOMEA terms contribute a significant portion of the time, as do the integral transformation, calculation of the MP2 energy, and formation of the transformed Hamiltonian (the latter three account for $\sim 85\%$ of $t_{\textrm{Other}}$). Grid generation, pruning, and THC fitting together account for $<3\%$ of $t_{\textrm{Other}}$ and $\sim 0.5\%$ of the total time. Such small errors encountered demonstrate the applicability of this approximation to larger systems for performing excited state and electron attachment calculations with high relative accuracy and significantly improved efficiency. 

\section{Conclusions}
We have demonstrated that the LS-THC approximation, in particular the robust variant, is efficient for reducing the scaling of coupled cluster singles and doubles and related equation-of-motion coupled cluster methods. The LS-PTHC and rCP-DF (R-LS-THC-like) approximations to the PPL term were previously shown to significantly reduce the computational time while maintaining high accuracy for the ground state CCSD method.\cite{THC-PPL-2014,THC-PPL-2021} Here, we have shown that these approximations, when applied to PPL term evaluations for excited state and electron attachment calculations, are highly effective in terms of accuracy and performance. The R-LS-THC approximation, in particular, is found to be highly accurate for a wide range of THC grid sizes while providing $\sim 5\times$ speedup for EOMEE-CCSD calculations with a triple-$\zeta$ basis set.

While we recommend a ``safe" choice of grid tolerance parameter $\epsilon$ of 0.01, larger values seem to have potential for situations where less accuracy is required and a very small grid is advantageous. It should be noted that for higher $\epsilon$ (typically, $\epsilon > 0.1$), we encountered some instability in the EOMEE-CCSD solution and phenomena such as changes in root ordering which complicated the assignment of states relative to the DF-EOMEE-CCSD calculation. It is not clear how much of these complications are due to a less-accurate evaluation of the $\sigma$ vector compared to poor ground state references in the corresponding CCSD calculations. Techniques such as orbital weighting\cite{THC-PPL-2014} could address some of the problems when using small grids, particularly with large basis sets.

Overall, the errors for the evaluation of excited states as well as of electron attached states do not seem to be related to chemical structures of the size of the molecules. These methods, in particular THC-PPL-EOMEA-MBPT2, are also applicable to large systems as demonstrated by calculations on a single-stranded DNA model system. Further reductions in walltime would require factorization of terms beyond the particle-particle ladder.

\begin{acknowledgement}
All calculations were performed on the ManeFrame II and Maneframe III high-performance computing facilities at SMU. This work was supported by the Robert A. Welch foundation under grant no. N-2072-20210327 and in part by the National Science Foundation (award CHE-2143725).
\end{acknowledgement}

\begin{suppinfo}
Electronic supplementary information files are available free of charge at the publisher's website. The following file is included:
\begin{itemize}
\item Supporting\_Information.xlsx: wall-time measurements for calculation of excitation energy of alkane chains, vertical excitation energies for molecules in QUEST1, QUEST3, and QUEST4 data sets, vertical electron-attachment energies for NAB22 molecules and single-stranded DNA model system with EOM-EA-CCSD and EOM-EA-MP2 methods, error in excitation energy for acetaldehyde with various auxiliary basis sets using DF and THC approximations. 
\end{itemize}
\end{suppinfo}

\bibliography{achemso-demo}

\end{document}